# Influence of surfactant, particle size and dispersion medium on surface plasmon resonance of silver nanoparticles

Vikash Sharma[1, $], Divya Verma[2], Gunadhor Singh Okram[1, #]

[1]UGC-DAE Consortium for Scientific Research, University Campus, Khandwa Road, Indore 452001, Madhya Pradesh, India.

[2] School of Studies in Chemistry & Biochemistry, Vikram University, Ujjain-456010, Madhya Pradesh, India

Email: [#]okram@csr.res.in, [$]vikashphy1@gmail.com

Clear influence of particle size, surfactants and dispersion medium on surface plasmon resonance (SPR) features of Ag nanoparticles (NPs), synthesized in thermal decomposition method, in the broad range of ultraviolet (UV) radiation, critical for many potential applications such as a photocatalyst, UV-sensor and detector, has been demonstrated here. It involves adsorbate coverage, interparticle distance or agglomeration, surface charge density and solvent refractive index (μ). NP agglomeration and surface charge density in solvents of varying μ have been studied systematically through zeta-potential (ζ) and hydrodynamic diameter (HD) using dynamic light scattering (DLS). The main SPR feature found at 316 nm in 31.5 nm NPs shifts to 320 nm in 15.1 nm NPs. The peak at 320 nm in air shifts to 259, 261 and 277 nm in polar solvent methanol, deionized water and ethanol, respectively and to 255, 275 and 282 nm in non-polar solvent n-hexane, benzene and toluene, respectively. In general, the decrease in particle size and increase in μ of solvents show red-shift. Curiously, a number of peaks up to seven in these solvents that are attributed to charge-transfer mechanism and change in inter-particle interaction of the NPs turning from a single peak of SPR in air has been observed for the first time. A model for re-adjustment of Fermi level ($E_F$) of Ag NP and the highest occupied molecular orbital (HOMO) and lowest unoccupied molecular orbital (LUMO) to explain them has also been used. Moreover, the Drude model for shift in the position of SPR in these NPs is only applicable in non-polar solvents, not in polar solvents. Such novel features will be potential candidates for various applications.

**Introduction**

The collective oscillations of charged density between metal surface and dielectric matrix, known as surface plasmon resonance (SPR), have always attracted much attention in research due to their excellent sensitivity to many surface properties[1]. Particularly, in nanoparticles (NPs), these oscillations are localized or confined in a small region, known as localized SPR (LSPR), resulting in enhancement of the amplitude of electro-magnetic field and hence strong resonance effects. However, these are hindered if dielectric loss is large. Ag NPs get much attention in this context since they have lowest loss in the visible spectrum compared to other plasmonic materials, and hence less attenuation of SPR. In addition, their excellent catalytic properties, high electrical and thermal conductivity makes them promising candidate for many potential applications[2,3] such as plasmon-enhanced light harvesting and photocatalysis[1,2], chemical, biological sensing and optics-based sensing[3,4,9], surface-enhanced Raman spectroscopy[18] and nanoelectronics and biomedical applications[4,5,6,7,3,8,9,10].

For example, Shen *et al.* showed that Ag NPs dispersed in water can be used as ink for inkjet printer[8]. Sheldon and co-workers reported that noble metal nanostructures have plasmo-electric potential that can convert light into electrical energy[7]. Recently, Frazer reported that Ag NPs can be used for treatment of human immunodeficiency virus (HIV)[11]. It is also found that ion, salt and NPs of Ag can be utilized in numerous consumer products and medical devices, due to its excellent antibacterial action[12]. In fact, many physical and chemical parameters such as particle size, size distribution, shape, inter-particle interaction or agglomeration of NPs, dielectric matrix or adsorbent/s and solvent refractive index[13,14,15,16] can influence SPR. Size dependent modifications on SPR such as shifting, broadening and damping in Ag NPs are more visible below the mean free path of electrons (52 nm)[17]. Solution phase synthesis of NPs is more efficient in this and easier one to tune the particle size, shape, size distribution and dielectric environment using variety of surfactants. However, synthesis of metal NPs is very challenging due to their tendency to quick oxidation.

The SPR in thin film nanowires and NPs of Ag is widely reported[4,18,1,17,19,20,21]. Several reports exist towards long wavelength (red-shift)[13,14,1] as well as towards short wavelength (blue-shift)[1,22,23] along with different positions of plasmon modes. For example, Peng *et al.* showed blue-shifts in SPR, as size decreases from 20 to 12 nm and then turns over in red-shift near 12 nm in ensembles of monodispersed silver nanospheres stabilized with oleylamine (OA) ligands in hexane. Scholl *et al.* reported the blue-shift in main characteristic feature of SPR with decrease in particle size from 20 to 2 nm in single Ag nanoparticle[17]. The blue-shift and damping in SPR with decrease in films thickness is also reported[24]. Ding *et al.* theoretically calculated SPR peak positions near 3.7 eV (333 nm) and 7.5 eV (165 nm) using Monte Carlo simulation method by means of reflection electron energy loss spectrum. The prominent surface plasmon peak overwhelms the feeble bulk plasmon feature near 333 nm, but peak near 165 nm exhibits both surface-and bulk-excitation contribution. It is found that the bulk plasmon feature is overwhelmed by the intense surface plasmon feature in the measured spectrum peak at about 3.7 eV. Yoon *et al.* reported two absorption peaks at 440 and 580 nm in nanopatterned metal thin film[21]. The main SPR peaks for Ag nanowires of diameters of 30–35 nm and 20–22 nm is in the range of 370–372 and 365–366 nm, respectively[25,19]. Recently, Jang *et al.* reported main SPR around 361 nm in 15 nm nanowires[26]. They found blue-shift in SPR peak with decrease in diameter of nanowire from 28 nm to 15 nm. Wang *et al.* reported the blue-shift in Ag nanocubes with decrease in size of cube[22].

However, these investigations have not yet performed the possible tuning SPR peak feature of metal surface and dielectric matrix using several solvents. They have been implemented successfully in this work using Ag NPs of Scherrer size ranging from 15.1 to 33.4 nm, synthesized using silver acetate, oleylamine (OA), trioctylphosphine (TOP) and polyvinylpyrrolidone (PVP) in very a simple and reproducible thermal decomposition method. We therefore demonstrate here the modification in dielectric environment around particles that led to exhibition of multi-peak SPRs observed first time and to changes in stability of the NPs. Their dispersion properties such as zeta potential ($\zeta$), hydrodynamic diameter (HD), electrical conductivity and mobility in various solvents are studied using dynamic light scattering (DLS). UV-Visible spectroscopy is used to study effect of particle size, inter-particle interaction or agglomeration, dielectric environment or cappants and solvent refractive index on SPR. In contrast to blue-shift SPR in air, it is found to be red-shifted, damped and narrower with decrease in particle size. Asymmetry at low wavelength side in the main SPR peak, with decrease in particle size, is also observed. Similarly, increase in refractive index ($\mu$) of the solvent leads to red-shift in both polar and non-polar solvents. An attempt has also been made to assess the applicability of Drude model for shift in main SPR peak in polar and non-polar solvents.

**Experimental**

Typically, 1 ml preheated TOP (90 %, Sigma Aldrich) at 210 °C is added in a solution of 1.5 g silver acetate ($\geq$ 99 %, Alfa Aesar) and 10 ml OA (70 %, Sigma Aldrich), already degassed at 110 °C for 30 min. The resulting solution is further heated at 190 °C for 2 h under argon atmosphere and cooled to 30 °C, and centrifuged in n-hexane to extract and wash the NPs. The washing is performed three to four times for remove the excess OA and TOP, not bound on NPs surface. The particles dried at 60 °C in vacuum drier are used for various characterizations. This sample was coded as Ag1. Samples synthesized, with 3 ml, 5 ml and 10 ml of TOP, respectively, with other reaction conditions remaining the same, are coded Ag2, Ag3 and Ag4. Two additional samples are prepared in 8 ml and 12 ml of TOP *only* at 190 °C for 2 h under argon atmosphere; they are coded as Ag5 and Ag6, respectively. Another sample, coded as Ag7, is also synthesized in a mixture of 10 ml of OA and 0.25 gram of PVP with other heating conditions remaining the same. They were synthesized thus to see the effect of adsorbates and surfactants on SPR. Schematic of synthesis of NPs and preparation conditions are shown in figure 1 and table 1, respectively.

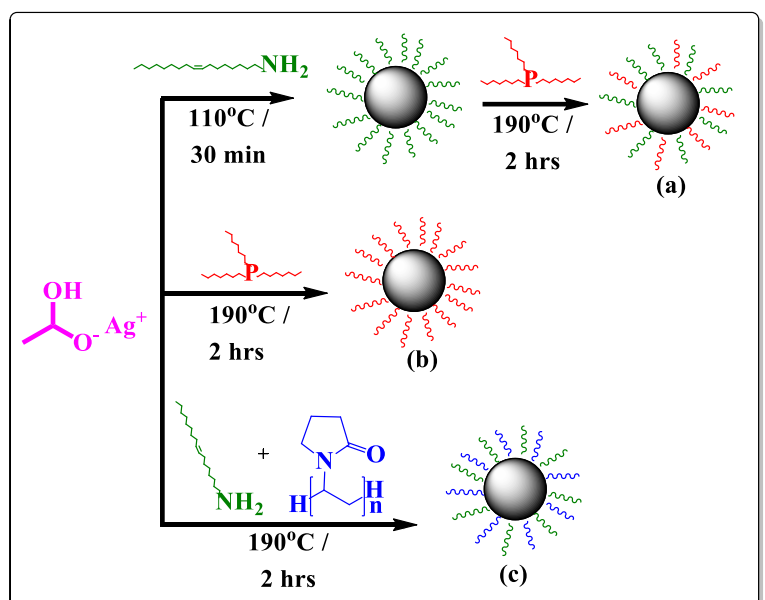

**Figure 1 Figure 1** Schematic representation of preparation and surface functionalization of Ag NPs using (a) oleylamine and trioctylphosphine, (b) trioctylphosphine only and (c) oleylamine and polyvinylpyrrolidone.

X-ray diffraction (XRD) data are collected using a D2 Phaser X-ray diffractometer with Cu K$_\alpha$ radiation (1.54 Å) in an angle range (2θ) from 20° to 90°. TEM and FESEM measurements are performed using TECHNAI-20-G$^2$ (200 KV) on drop-casting the well-sonicated dispersion of NPs in ethanol on carbon-coated TEM grids and FEI Nova nanosem450, respectively. To confirm that coating on the surface of NPs, we did Fourier transformed infrared (FTIR) spectroscopy using an FTIR spectrometer (Bruker, Vertex 70), and valence state is examined using X-ray photoelectron spectroscopy (Al-Kα, λ = 0.834 nm). The UV-Visible data in diffused reflectance mode are collected using Perkin Elmer Lambda 950 in the wavelength range 200 nm to 800 nm with resolution of 0.05 nm and wavelength accuracy of ±0.08 nm. Particle size and zeta potential (ζ) measurements using a dynamic light scattering (DLS)-based Zeta/NP analyser (NanoPlus-3) are performed after thorough sonication of the NPs dispersed in various solvents viz., methanol, deionized water, ethanol, n-hexane, benzene and toluene to study the effect of dielectric environment with different refractive indices and stability of particles. Concentration of NPs added is 0.1 mg/ml in each case.

**Table 1** Sample synthesis conditions, particle size and SPR peak positions (in air)

| Sample | TOP (ml) | OA (ml) | PVP (gram) | XRD size (nm) | TEM size (nm) | SPR peak position (nm) |
|---|---|---|---|---|---|---|
| **Ag1** | 1 | 10 | - | 31.5 | - | 316 |
| **Ag2** | 3 | 10 | - | 28.9 | 29.1±1.2 | 318 |
| **Ag3** | 5 | 10 | - | 24.7 | 25.3±0.9 | 318 |
| **Ag4** | 10 | 10 | - | 15.1 | 15.9±1.1 | 320 |
| **Ag5** | 8 | - | - | 29.6 | 60±2 | 318 |
| **Ag6** | 12 | - | - | 24.6 | - | 320 |
| **Ag7** | - | 10 | 0.25 | 33.4 | 41.7±1.2 | 318 |

## Result and Discussion
### X-ray diffraction

XRD patterns of Ag1, Ag2, Ag3, Ag4, Ag5, Ag6 and Ag7 are shown in figure 2. Ag1 exhibits peak positions at 2θ = 38. 40$^0$, 44.20$^0$, 64.36$^0$, 77.33$^0$, 81.47$^0$ due to reflections from (111), (200), (220), (311) and (222) planes, respectively. They are confirmed to be face centered cubic (fcc) phase of Ag (JCPDS# 893722) without any impurity peak. Similar results are evident from the XRD patterns of other remaining samples indicating that their respective crystal structure is fcc phase without any impurity phase. Notably, the peaks are broadened with increase in concentration of TOP, which indicates reduction in particle size. The average particle size of NP samples is evaluated using Scherrer's formula (table 1).

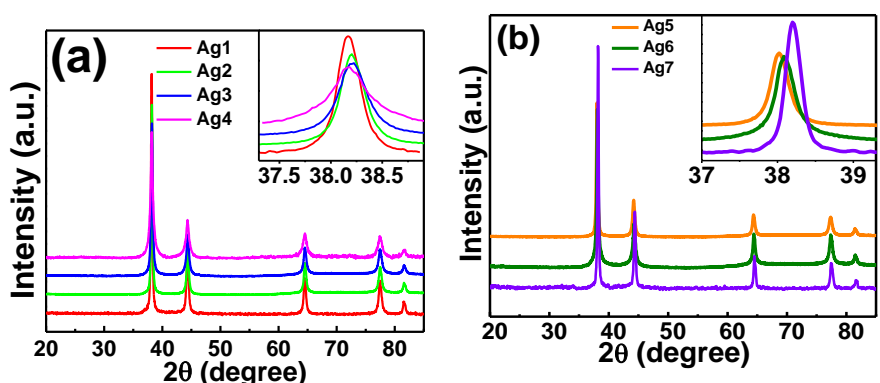

**Figure 2** X-ray diffraction patterns (a) Ag1, Ag2, Ag3 and Ag4, and (b) Ag5, Ag6 and Ag7.

### EDX and XPS study

Figure 3 shows the EDX spectra of Ag4, Ag5 and Ag7 that show peak related to Ag without any impurity peak. Quantitatively, 100 % Ag content is found, since others due to surfactants are undetectable. To see the valence state or any oxidation of NPs, XPS measurements are performed on Ag4. Fitted spectra using XPSPEAK4.1 software of C 1s, O 1s and Ag 3d along survey scanned are shown in figure 4. The two peaks near at 285.1 eV and 286.7 eV in C 1s, correspond to C-C and C-O, respectively.

Two peaks around 531.8 eV and 532.9 eV of O 1s attributed to low coordination bonding with -OH and C-O-H or -CO$_3$, respectively related to surfactant/s or adsorbed hydrocarbon species, not due to presence of Ag$^{+1}$ or Ag$^{+2}$ states since oxides generally narrower peak width and give rise to two well-resolved peaks in O 1s[27]. Furthermore, two well resolved peaks around 368.3 eV and 374.3 eV in Ag 3d corresponding to spin-orbital components, Ag $3d_{5/2}$ and Ag $3d_{3/2}$, respectively, are observed. The peak around 368.3 eV and the separation between spin-orbital components found to be 6.0 eV are evidence for metallic Ag[27]. Therefore, EDX and XPS results confirm the single phase of metallic Ag NPs and therefore oxide of Ag is ruled out.

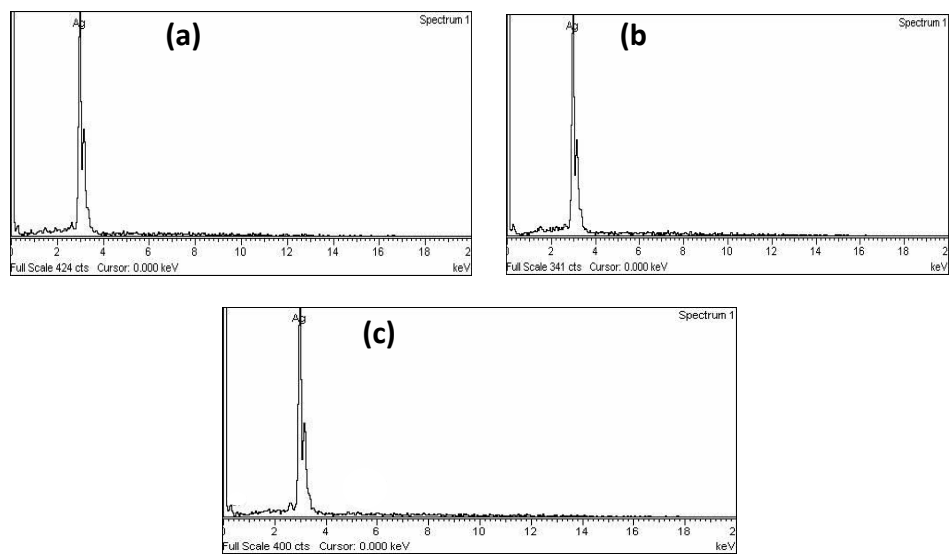

**Figure 3** Energy dispersive x-ray spectra of (a) Ag4, (b) Ag5 and (c) Ag7.

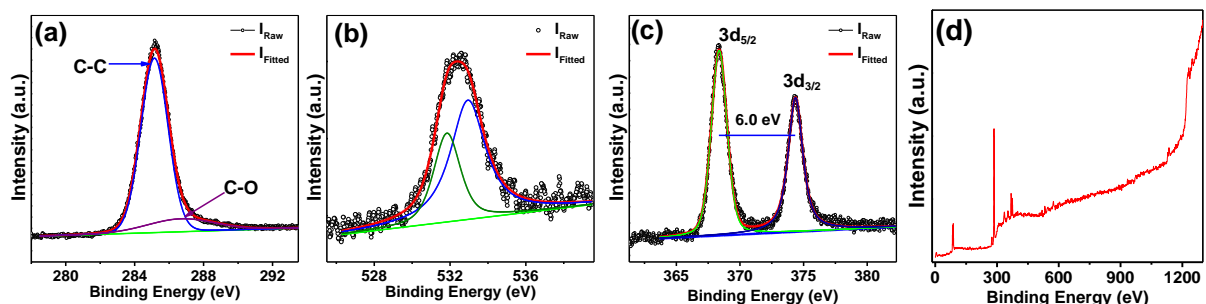

**Figure 4** XPS spectra of (a) C 1 s, (b) O 1s, (c) Ag 3d and (d) survey scan of Ag4.

**HRTEM and FESEM study**
The TEM micrographs and their size distributions of Ag2, Ag3, Ag4, Ag5 and Ag7 are shown in figure 5 and HRTEM of Ag2 is shown in figure 6. The size distribution of NPs is found to be between 10 to 40 nm, 10 to 40 nm and 10 to 20 nm with average particle size around 29.1±1.2 nm, 25.3±0.9 nm, 15.9±1.1nm in Ag2, Ag3 and Ag4 as shown in figure 5 (d, e, f), respectively. The decrease in particle size and enhanced monodispersity of NPs with increase in TOP are found such as NPs of Ag4 are reasonably monodispersed (figure 5 (c)) compared to Ag2 (figure 5 (a)). The average particle size is ~ 60 nm for Ag5 (figure 5 h). Therefore, NPs prepared in pure TOP seem to be more agglomerated and bigger in size compared to that prepared in combination of OA and TOP (figure 5 d).
The shape of NPs prepared in TOP and combination of OA and TOP appears generally spherical. The particle size of NPs obtained from XRD and TEM are nearly the same in Ag2, Ag3 and Ag4 (table 1). TEM size of Ag5 (60 nm) found to be significantly larger than XRD size due to agglomeration of smaller NPs (table 1). This is attributed to the smaller crystallites (XRD size) being agglomerated into bigger size particle in TEM. Similarly, very random triangular, rectangular, hexagonal and spherical shapes of NPs are found in Ag7 (figure 5 (i, j)) with an average size of about 41.7 nm, which is significantly larger than the XRD size (Table 1). Finally, the selected area electron diffraction (SAED) patterns of these samples, shown in figure 5, inset (a, b, c, d, i), confirm the fcc structure seen in XRD (figure 2).
High resolution TEM data for Ag2 are shown in figure 6 (a-c). SAED patterns clearly indicates the reflection from (111), (200), (220) and (311) planes of fcc phase of Ag NPs (figure 6 (d)). The interplanar spacing (d) around 0.233 nm and 0.215 nm corresponds to (111) and (200) planes, respectively (figure 6 (b, c). The fast Fourier transformed (FFT) patterns of these planes are shown in figure 6 (e, f). These results prove that Ag2 is fcc Ag NPs.

FESEM micrographs of Ag4 in powder form and after dispersing in ethanol or n-hexane and consequent drop-casting on glass slides are shown in figure 7. The measurements on dispersed NPs have been performed to see the effect on morphology and inter-particle interactions or agglomeration in comparison with powder sample. Morphology of NPs found to be hexagonal-like in powder form as well as in n-hexane and ethanol. They look to be more agglomerated in n-hexane 7 (c, d) compared to powder (figure 7 (a, b)). This means that inter-particle distance is decreased in n-hexane. Furthermore, inter-particle interaction or agglomeration is less in ethanol compared to n-hexane (figure 7 (c, e)). Ring in FESEM micrographs (figure 7 c, e) indicates the agglomerated clusters of NPs. They appear relatively distinctly different from that in air or the powder form. Notably, the identical morphology of NPs before and after dispersing in ethanol and n-hexane are noted. The increase in agglomeration of NPs or decrease in inter-particle distance may modify their spatial charge distribution and hence in their SPR, to be discussed later.

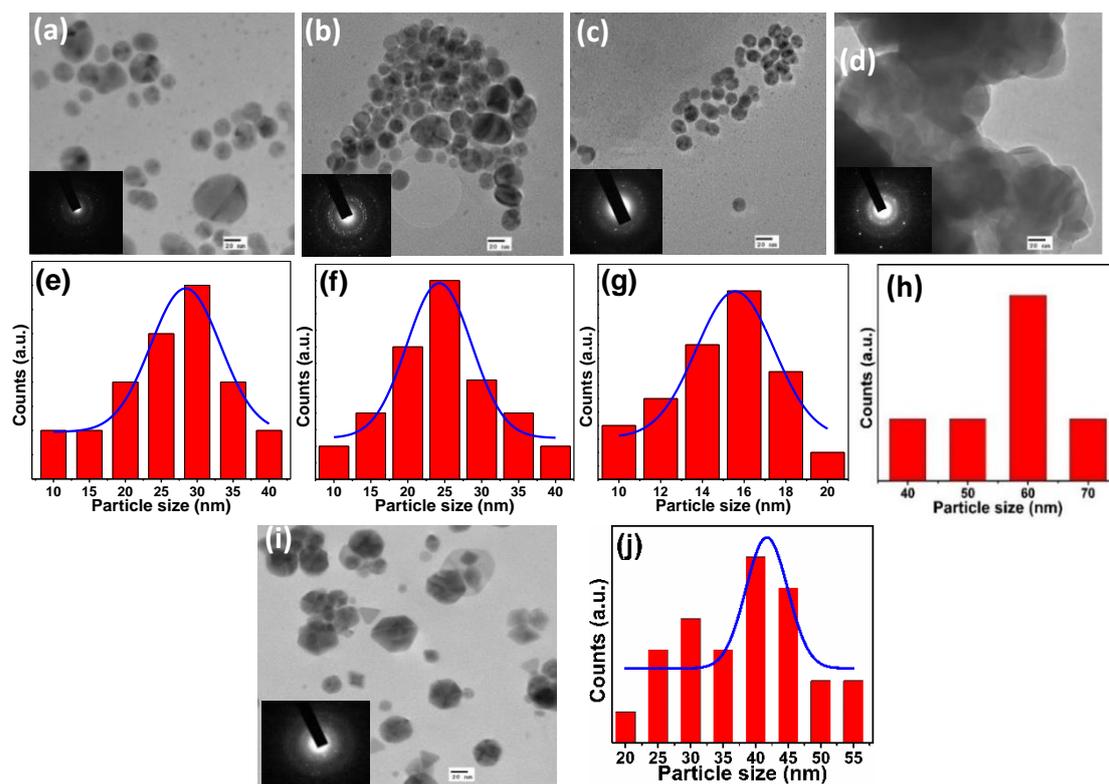

**Figure 5** TEM micrographs (a, b, c, d, i) and size distribution (e, f, g, h, j) of Ag2 (a, e), Ag3 (b, f), Ag4 (c, g), Ag5 (d, h). and Ag7 (i, j). Scale bar of all images is 20 nm.

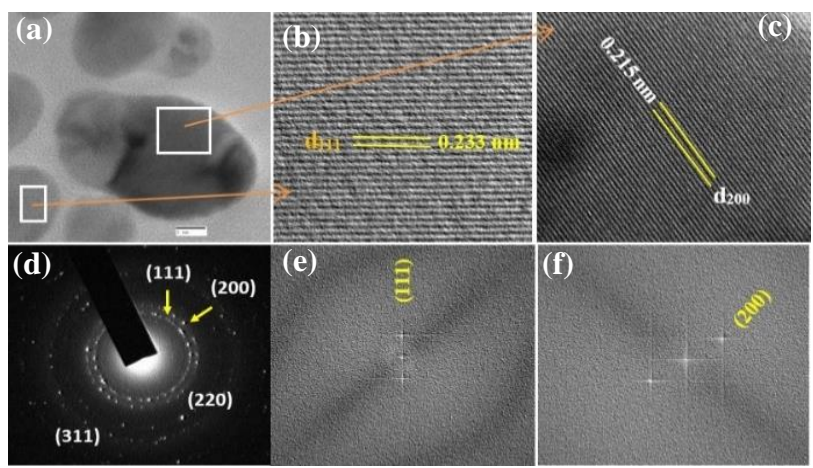

**Figure 6** HRTEM micrographs (a, b, c), (d) SAED and (e, f) FFT of plane (111) and (200), respectively of Ag2. Scale bar of (a) is for 5 nm.

**FTIR study**

FTIR spectra of Ag1, Ag5 and Ag7 are shown in figure 8. A broad absorption peak near 3425 cm$^{-1}$, 3440 cm$^{-1}$ and 3442 cm$^{-1}$ is assigned to –OH stretching vibration in Ag1, Ag5 and Ag7. The stretching modes of $CH_2$ at 2924 cm$^{-1}$ and 2848 cm$^{-1}$ in Ag1, 2924 cm$^{-1}$ and 2852 cm$^{-1}$ in Ag5 and 2921 cm$^{-1}$ and 2854 cm$^{-1}$ in Ag7 are seen. The peaks near 1220 cm$^{-1}$ and 1034 cm$^{-1}$ are present in Ag1 mainly due to C–P and C–N stretching modes of TOP and OA[28,29], which confirms the presence of OA and TOP on surface of Ag1. Peaks near 1070 cm$^{-1}$ and 1210 cm$^{-1}$ in Ag5 are attributed C–P modes of TOP. Furthermore, a sharp peak at 671 cm$^{-1}$ in TOP coated NPs (Ag5) assigned to =C-H bending or $(-CH_2-)_n$ ($n \geq 4$) stretching of TOP[30,31]. The peaks near 1661 cm$^{-1}$ and 880 cm$^{-1}$ are due to -C=O and pyrrolidone ring breathing vibration, respectively, of PVP, and peaks between 1026-1076 cm$^{-1}$ and 1291 cm$^{-1}$ are mainly due to -C-N vibration of OA and PVP[32,33]. Additional peaks between 1424 cm$^{-1}$ is due to -$CH_2$ bending vibration of PVP. These results indicate that Ag1, Ag5 and Ag7 are coated with OA-TOP, TOP and OA-PVP, respectively.

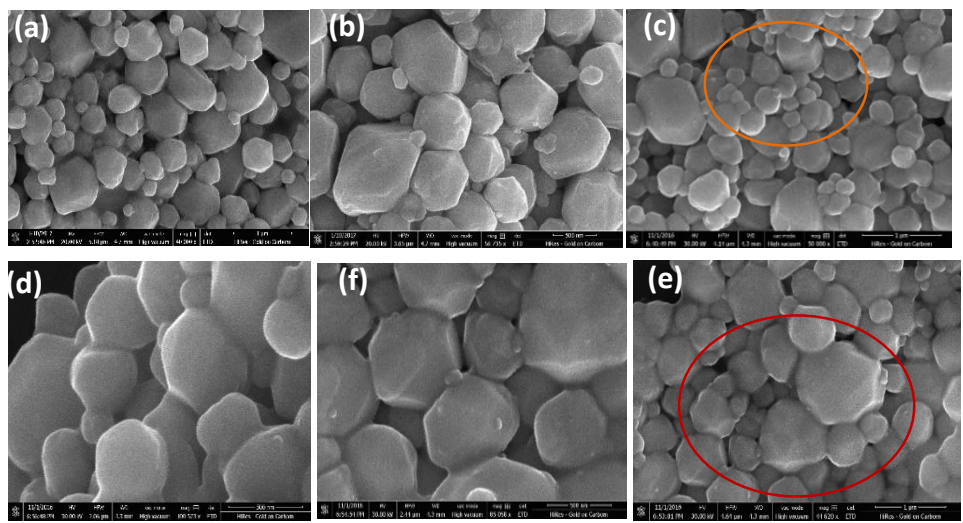

**Figure 7** FESEM micrographs at different scales in (a, b) powder form on carbon tape, (c, d) dispersed in n-hexane and (e, f) ethanol and drop-casted on glass slides of Ag4. The scale bar is 1μm in (a, c, e) and 500 nm in (b, d, f).

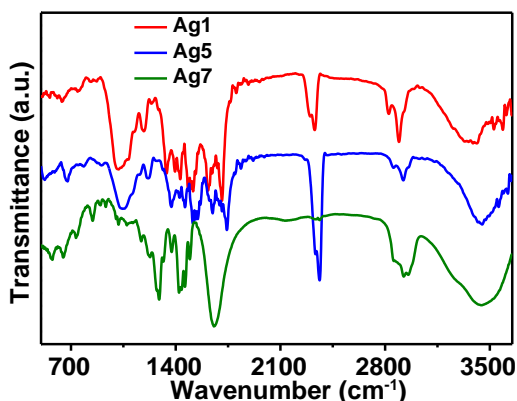

**Figure 8** Fourier transformed infrared spectra of Ag1, Ag5 and Ag7.

**Dynamic Light Scattering Study**

Intensity distribution of HD of Ag4 in different solvents at pH=7 are shown in figure 9 and listed in table S1. HDs in methanol in the first run 250 and 25930 nm turn 257 nm and 26730 nm in the second run. They are 145 nm & 13165 nm and 290 nm & 13970 nm in DIW, 160 nm & 13450 nm and 170 nm & 15860 nm in ethanol, 38 nm, 200 nm & 6820 nm and 42 nm & 2752 nm in n-hexane, respectively. However, in benzene and toluene, there is a single peak each at 12210 nm and 15080 nm, respectively. Notably, while two peaks each in methanol, DIW and ethanol in both runs are found with increased values in the second run, there are three peaks in the first run and two peaks in the second run in n-hexane with increased value in the smallest value but reduced value of the bigger value in the second run. Such results are very intriguing since dispersion characteristics of a particular nanoparticle is so divergent depending on the type of dispersant such as these. They will influence significantly the zeta potential values and SPR behaviour of the NPs in these dispersant as presented below.

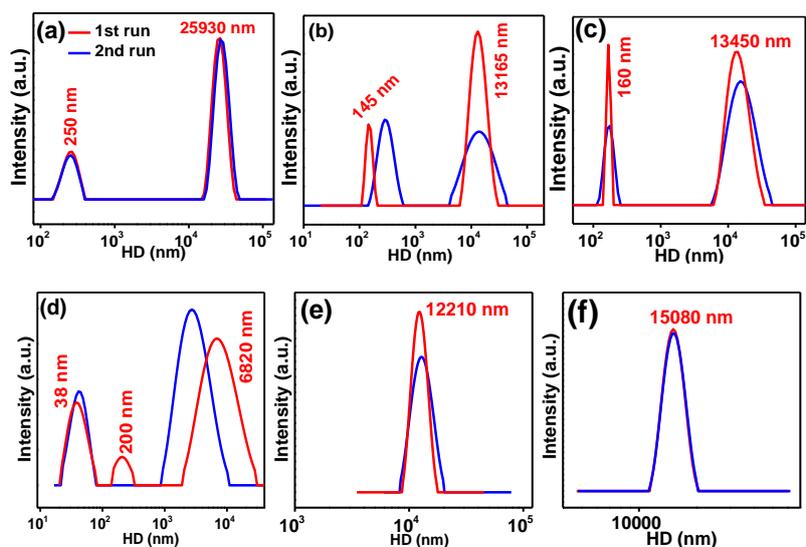

**Figure 9** Intensity distribution of hydrodynamic diameter of Ag4 in (a) methanol, (b) deionized water, (c) ethanol, (d) n-hexane, (e) benzene and (f) toluene. The red and blue curves are first run and second run, respectively as shown in (a).

Figure 10 shows the intensity distribution of ζ of Ag4 in various solvents at pH=7. It is interesting to note that five peak values of ζ at 0.9 mV, 22.70 mV, 14.60 mV, 33.60 mV and 82.70 mV in the first run turn to be at 13.45 mV, 26.80 mV and 74.50 mV in the second run in methanol. Two peaks at -37.23 mV & -32.30 mV in the first run remains to be two peaks at -36.02 mV and -28.30 mV in the second run in DIW. Two peaks at 10.40 mV and -171 mV turn to be two peaks at 9.80 mV and -147 mV in the second run in ethanol. Four peaks at 89.80 mV, 26.70 mV, 26.94 mV and -375 mV in the first run became three peaks at 67.90 mV, 22.50 mV and 345 mV in the second run in n-hexane. One peak at 8.70 mV in the first run remains to be one peak at 7.16 mV in benzene. Similarly, in toluene, the single peaks are at 5.10 mV and 3.40 mV, respectively. These data are therefore very fascinating as they reflect very complex nature of the presence of different kinds of agglomerated particles which are not distinguishable in HDs. This would be the reason why even though only two peaks are seen in both the runs in HDs, there are five peaks in the first run but three peaks are seen in the second of the ζ plot for methanol. Complexity reduces in n-hexane. Similarly, in DIW and ethanol with two peaks each and to the least in benzene and toluene with a single peak each. The lager ζ corresponds to smaller particles and smaller ζ corresponds to agglomerated clusters as found in HD data. Interestingly, only one peak in ζ in benzene and toluene is found, which is consistence with their one size distribution peak in HDs. Also, two well-separated peaks in methanol, ethanol, DIW and n-hexane can be seen in ζ. Moreover, the splitting in main peak in methanol, DIW and n-hexane can clearly be seen, which is more in methanol and n-hexane compared to DIW. Notably, NPs exhibit both negative as well as positive charges or ions within the electrical double layer in ethanol and n-hexane, whereas only positive ions in methanol, benzene, toluene and negative ions in DIW are detected. The ζ, HD, mobility and conductivity of ions obtained from DLS measurements are shown in table S1. Generally, Ag NPs are found to be more stable in DIW (ζ ≥ ± 30 mV) compared to other solvents.

The average ζ around 30.4 mV, 25.2 mV, 18.6 mV and 9.4 mV for Ag1, Ag2, Ag3 and Ag4, respectively, in ethanol are found (figure 11 (a)). The value of ζ decreases as the size decreases, which is illustrated in figure 11 (b). The positive value of ζ for Ag1 in ethanol is line with OA and TOP coated Ni NPs[28], and decrease in ζ with increase in quantity of TOP as a medium. The ζ highly depends on the surface chemistry or charge density and functional group per surface area of NPs in comparison of particle size, which is line with earlier report on $SiO_2$ NPs[34]. This might probably due to increase in concentration of TOP or increase in phosphine groups on surface of NPs, consequently neutralization of positively surface charges in electrical double layer as seen in Ag1. Therefore, the increase in agglomeration or decrease in zeta-potential is mainly due to increase in coverage of TOP or its thickness around NPs owing to the modified chemical interactions and surface charge density rather than decrease in particle size (figure 11 (b)). Narrower size distribution of ζ in Ag4 than other samples further support its relatively more monodispersity as seen in TEM micrographs (figure 5 (a, b, c)).

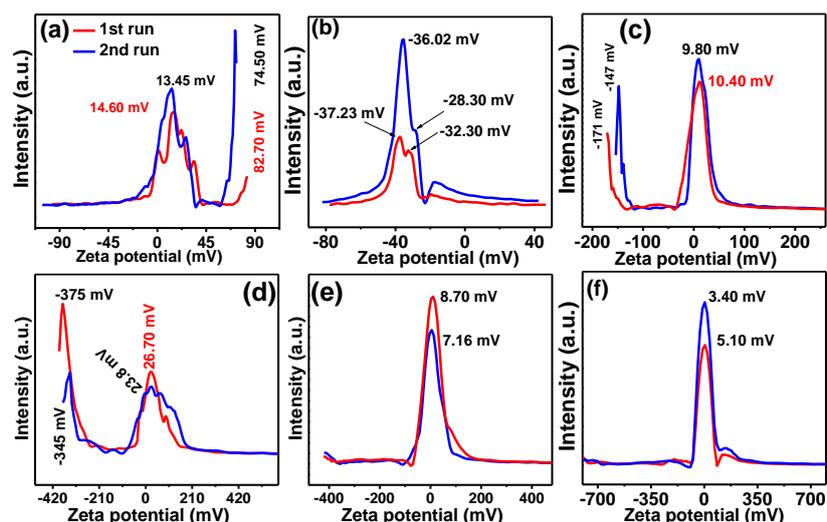

**Figure 10** Intensity distribution of zeta-potential of Ag4 dispersed in (a) methanol, (b) deionized water, (c) ethanol, (d) n-hexane, (e) benzene and (f) toluene. The red and blue curves are first run and second run, respectively as shown in (a).

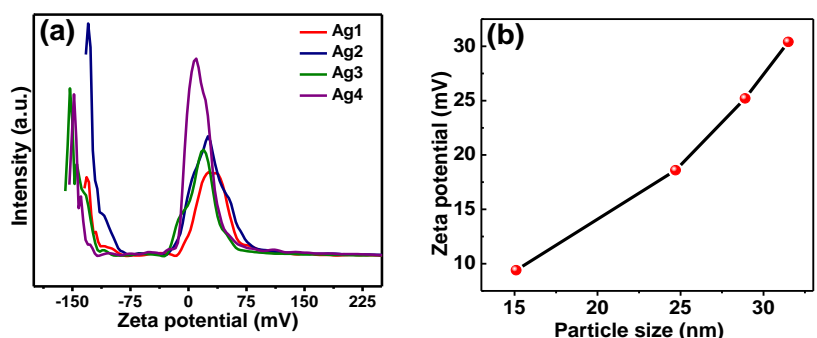

**Figure 11** (a) Zeta potential of Ag1, Ag2, Ag3 and Ag4, and (b) size dependence on zeta potential in ethanol.

**Influence of particle size and adsorbate coverage on UV-Vis spectroscopic response**

The diffused reflectance spectra of Ag1, Ag2, Ag3, Ag4, Ag5, Ag6 and Ag7 are shown in figure 12. Dips found to be near 220, 320, 378, 454, 498 and 554 nm in Ag4 are denoted by P1, P2, P3, P4, P5 and P6, respectively (figure 12 (d)). They appear at 220, 316, 378, 488 and 558 nm in Ag1 (figure 12 (a)), 220, 318, 378, 488 and 558 nm in Ag2 (figure 12 (b)), 220, 318, 378, 490 and 556 nm in Ag3 (figure 12 (c)) and 220, 318, 374, 484 and 562 nm in Ag5 (figure 12 (e)), 220, 320, 375, 450, 498 and 558 nm in Ag6 (figure 12 (f)) and 218, 318, 378 and 500 nm in Ag7 (figure 12 (g)). Peak (P1) found to be nearly at same position in Ag1, Ag2, Ag3, Ag4, Ag5 and Ag6, is attributed to inter-band transition (IBT) in metal NPs[16]. It however shifts to lower wavelength side (blue-shift) by 2 nm in Ag7, which might be due to different dielectric environment or shape of NPs rather than particle size effect, as it doesn't show size dependence in other samples. Furthermore, P2, P3, P4, P5 and P6 are assigned to SPR in Ag NPs, consistent with earlier reports[1,4,18,17,19,20,21] on thin films and NPs of Ag with varying grain sizes, different shapes and dielectric environment. Appearance of P2 is most likely due to dipole SPR, while others are due to quadrupolar and higher multipolar plasmon modes, mainly due to imperfect spherical shape and coupling between dipolar modes of SPR in adjacent NPs[35]. This finding of SPR in UV-regime in Ag NPs is closed to the bulk plasmon[4]. Nevertheless, most intense and its extended features with variation in particle size, dielectric environment and solvent refractive index, confirmed the dipole SPR, consistent with earlier reports[1,4,18,17,19,20,21].

The main SPR peak (P2) found to be red-shifted from 316 to 320 nm with decrease (increase) in particle size (TOP), contrary to blue-shift, is in line with earlier report[1]. Significant asymmetry of P2, at the lower wavelength side, increases with increase in particle size, which may be due to broad size distribution of NPs, as seen TEM data (figure 5). Plasmon modes of NPs may interfere resulting in asymmetry in peak[14,36]. Furthermore, P2 and P5 are red-shifted in Ag7 compared to Ag1, contrary to blue-shift. Moreover, P6 is completely disappeared in Ag7. These results may due to different shape, size distribution and dielectric environment rather than particle size. As particle size decreases, P2 and P5 found to be red-shifted, whereas P6 is blue-shifted in Ag1, Ag2, Ag3 and Ag4. Moreover, while P3 doesn't show any size dependence, P4 is present in Ag4 and Ag6 only. Similarly, P2 and P5 are red-shifted and P6 is blue-shifted, while P3 did not shift with decrease in particle size. These results are consistent with earlier reports of red-shift[1,14] and blue-shift[23] in plasmon modes for Ag NPs with decrease in particle size. Furthermore, disappearance of P4 with increase in particle size might be due to coupling between plasmonic features, their relevance increases as size decreases, which leads to new feature in smaller NPs[35].

Now, there are two probable situations: first, increase in concentration of TOP, which favour in red-shift and second, decrease in particle size favours of blue-shift. Generally, decrease in particle size in bare NPs may have weaker interactions of s-d electrons due to finite size and surface effects, which may lead to increase in plasmon energy or blue shift in SPR features[24]. Therefore, the unusual redshift is most likely to be due to effect of TOP coverage or increase thickness of dielectric matrix, consistent with earlier report[37]. To verify this, UV-Visible measurements were performed on Ag1, Ag2, Ag3 and Ag4 in ethanol (figure 13). There is clear increase in red-shift P2 with increase in TOP concentration. On the contrary, P2 is blue-shifted to near or less than 275 nm and split in three peaks in ethanol compared to air 316-320 nm (figure 12 (i)) as illustrated in inset of figure 13 that will be discussed later further. It is gradually broadened, damped and become asymmetric as the particle size decreases, consistent with earlier report[36], and our reflectance data (figure 12). Decrease in absolute ζ or reduce surface charge density and inter-particle distance or increase in agglomeration with decrease in particle size due to more uniform size compared to those of larger particle size samples as reported earlier[28] or as discussed above also corroborate red-shift, consistent with earlier reports[1,37].

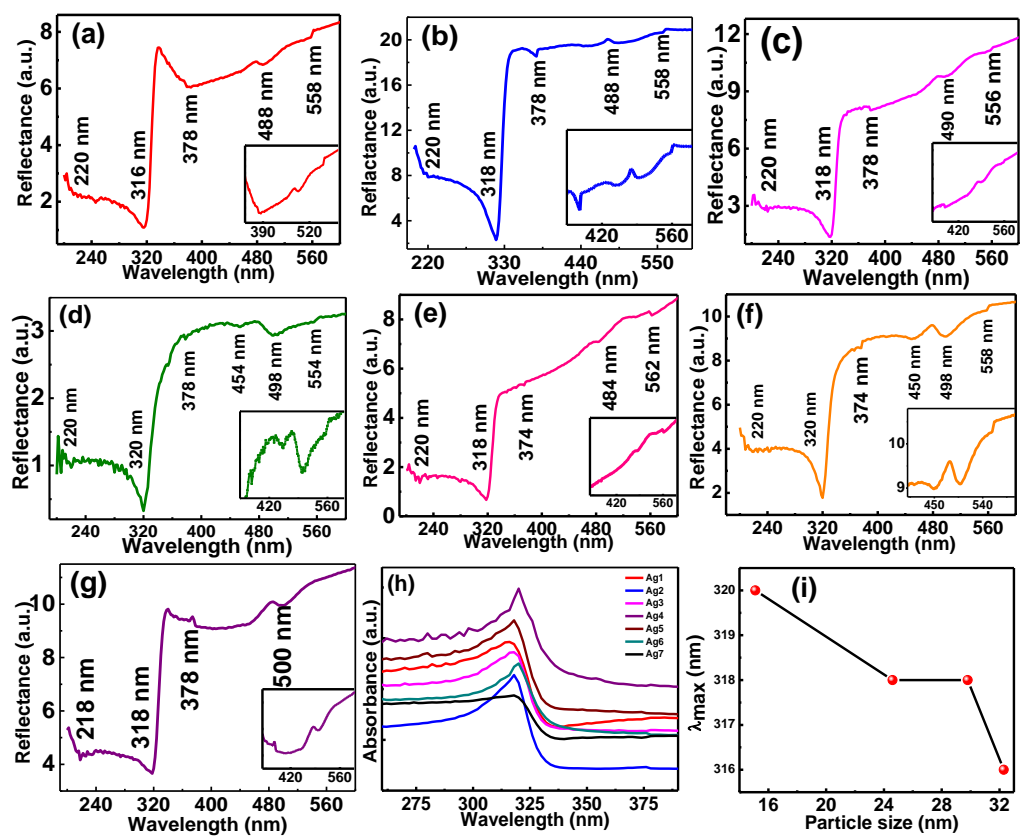

**Figure 12** Diffused reflectance spectra in air of (a) Ag1, (b) Ag2, (c) Ag3, (d) Ag4, (e) Ag5, (f) Ag6, (g) Ag7, (h) corresponding absorbance spectra of these samples and (i) wavelength at peak $\lambda_{max}$ versus particle size.

**Influence of refractive index of solvent on SPR**

In order to find the possible origin of peak splitting and effect of solvent refractive index on SPR, as one peak found in reflectance data in air (figure 12) splits into three independent peaks (figure 13), UV-Visible absorbance spectra of Ag4 in several polar solvents, viz., methanol ($\mu$= 1.327), deionized water ($\mu$= 1.3325) and ethanol ($\mu$= 1.3617), and non-polar solvents, viz., n-hexane ($\mu$=1.3758), benzene ($\mu$= 1.4957) and toluene ($\mu$= 1.4964) are recorded (figure 14). It is noted here that same dispersion of NPs is used for these measurements as those of DLS and FESEM for checking the consistency of the data. Peaks around at 214 and 259 nm in methanol, 261 nm in DIW, 221 and 279 nm in ethanol, 209 and 255 nm in n-hexane, 216 and 275 nm in benzene and 216 and 282 nm in toluene are observed. Position of the SPR peak in each sample is determined using Gaussian fitting (figure S1). The first (P1) and second (P2) peaks are related to IBT and SPR, respectively, with modified shape and shifting as found in air. However, remaining peaks (P3, P4, P5 and P6 in figure 12) are not visible in these solvents due to their probable damping with increase in $\mu$[16,38]. Systematically red-shift in IBT and SPR in both polar as well as non-polar solvent with increase in $\mu$ is found, consistent with earlier report[15]. Interestingly, IBT is disappeared in DIW. The SPR is significantly damped and weaker in benzene and toluene, probably due to their higher µ. To see the accuracy of the data, measurements are repeated in n-hexane and benzene; they encouragingly show identical features.

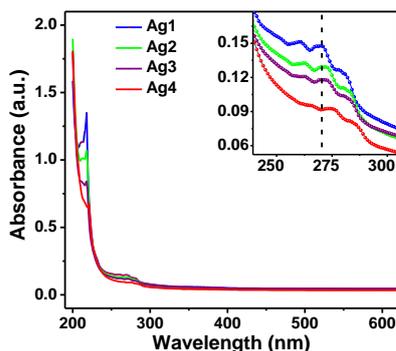

**Figure 13** UV-Visible spectra of Ag1, Ag2, Ag3 and Ag4 in ethanol.

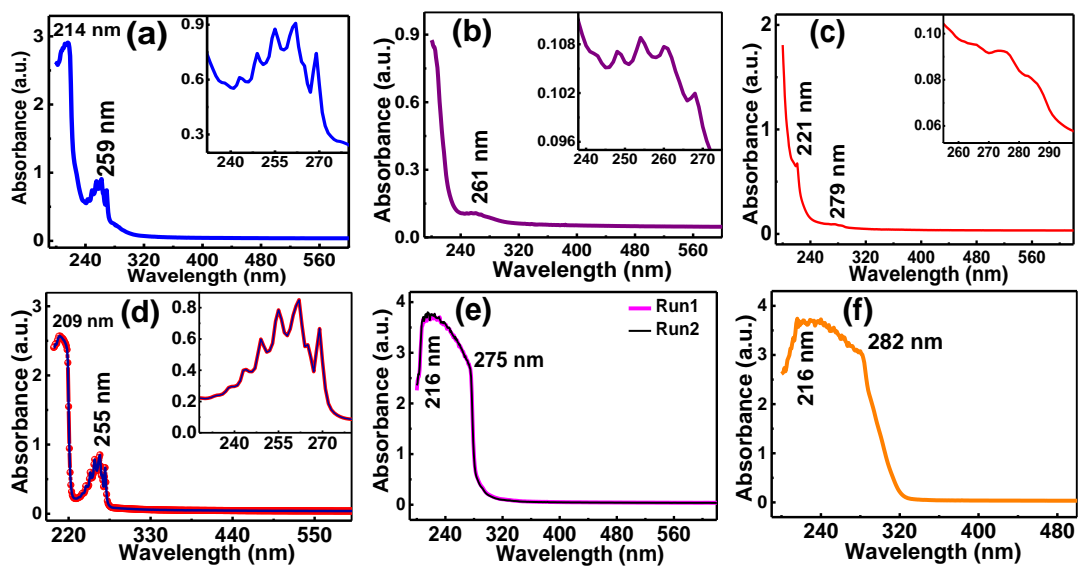

**Figure 14** UV-Visible absorbance spectra of Ag4 after dispersed in (a) methanol, (b) DIW, (c) ethanol, (d) n-hexane, (e) benzene and (f) toluene.

Remarkably, splitting into three, four, five and seven peaks in ethanol, DIW, methanol and n-hexane, respectively, for a single SPR peak exhibited in air has been observed for the first time. They therefore are distinctly different from that observed in Ni NPs[16]. This may be understood by charge transfer mechanism and formation of interface dipole between NPs and surface-bound molecules of the solvent or the adsorbate. When the NPs are dispersed in any medium, the molecules of solvent may adsorb on metal NPs either by physisorption or chemisorption. Consequently, electronic structure of NPs and adsorbate may be modified. Further, charge transfer may take place between NP and adsorbate, if they are chemisorbed, which may affect interface dipoles. The exchange of electrons through chemical bonds will highly depend upon their relative electronegativities.

Metal work function ($W_m$) may be suppressed in this due to repulsion between electrons of molecule and metal surface electrons and the direction of the dipole moment will be towards vacuum level (VL) of adsorbate [39]. This further result in abrupt shift in vacuum level from metal to adsorbate molecule interface and hence formation of interface dipole barrier. This means that the energy difference between the Fermi level ($E_F$) of metal NP and the highest occupied molecular orbital (HOMO) and lowest unoccupied molecular orbital (LUMO) of the molecule may change with respect to a vacuum level alignment situation and charge transfer will depend upon direction and magnitude of interface dipole barrier height. The combined positive pole of interface dipole pointing towards metal NP and negative pole towards surface-bound molecule lead to decrease in the Fermi energy of metal NP and increased HOMO energy of surface-bound molecule by the addition of electrostatic energy and vice-versa[40]. The schematic of negatively and positively charged metal NPs along with formation of interfacial dipole is shown in figure 15.

When the surfactant molecules are in contact with metal NP, their electronic energy levels may be modified due to their different work functions or vacuum level (VL). For negatively charged metal NPs, work function may be suppressed with direction of dipole towards VL of adsorbate (figure 15 (a)). The direction of dipole will revert back for positively charged surface of NPs than that of negatively charged surface (figure 15 (b)). Therefore, strength of injection barriers for electrons ($\Delta_e$) and holes ($\Delta_e$) may be modified according to direction and height of dipole barrier ($V_{dipole}$) and hence charge transfer; electronic energy level of NPs are taken discrete due to finite size effects. Schematic representation of interaction of OA-TOP coated NPs with chemically attached molecules of solvent (upper panel) and wavevector of surface plasmon polariton (SPP) or SPR for single NP and ensemble of NPs (lower panel) is shown in figure 15 (c). From this we can say that any change in interparticle distance and arrangement of agglomerated particles highly affect the SPR features[14]. The interaction between negatively charged (figure 15 (d)) and positively-charged (figure 15 (e)) metal NPs with the medium; corresponding to direction of dipole towards (figure 15 (a)) and away from (figure 15 (b)) of VL of adsorbate, respectively.

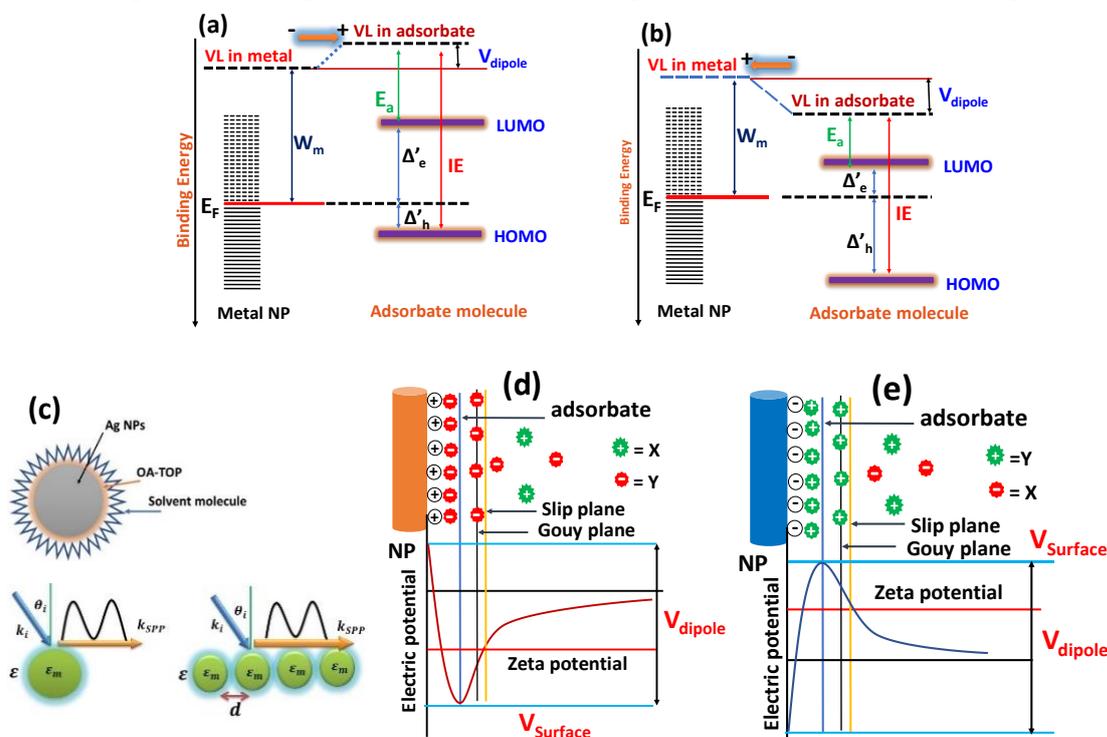

**Figure** 15 Electronic energy levels at metal/adsorbate interface (a) interface dipole directed towards adsorbate and (b) interface dipole directed towards metal, due to vacuum level (VL) mismatch. (c) Upper panel: Interaction of NP-coated with OA-TOP with solvent molecules and Lower panel: effect of interparticle interaction or agglomeration of NPs on surface plasmon polariton. (d) Negatively-charged and (e) positively-charged metal NPs in the medium corresponding to (a) and (b), respectively.

In this context, LUMO ~3.50 eV of methanol is closed to Fermi energy ~ 5.49 eV of Ag compared to HOMO~-11.14 eV. While, HOMO ~ 5.39 eV of ethanol is much closed to Fermi level of Ag than LUMO ~ -2.42 eV. Methanol has less electrons injection barrier than ethanol and ethanol have lesser hole injection barrier than methanol. Therefore, charge transfer nature will be changed for different solvents. HOMO and LUMO of adsorbates i.e. OA and TOP are not considered for simplicity, they however will also involve in this charge transfer mechanism. Hence, the combination of all means metal/surfactant and solvent give such type of features, which is difficult to rule out here[41]. P2 is split or oscillate in methanol, ethanol, DIW and n-hexane in three, four, five and seven, respectively. The number of split peaks is different in these solvents, which may be understood through their different charge-transfer nature, electronegativity, chain length and coverage of molecules, and hence surface charge density or interface dipole[40].

The transferred charge or formed dipoles may oscillate with slightly different frequency than that in air, which leads to splitting in SPR. Furthermore, the higher ζ is related to larger surface charge density or more dipoles, which may result in more oscillations in P2 in n-hexane than other solvents[40]. In addition, larger blue-shift in P2 in n-hexane compared to ethanol is observed while they have nearly equal value of μ, which is due to different availability of number of electrons[36]. Less ζ and larger HD in ethanol compared to n-hexane indicates more agglomeration of NPs or decrease in interparticle distance, which is consistent with FESEM micrographs (figure 7). As a consequences, red-shift and new peak may appear in optical spectra compared to a single or unagglomerated NP, due to coupling of plasmon modes in adjacent NPs[35]. The larger broadening, damping and asymmetry in ethanol support this picture[14].

The variation in position of SPR with µ may be explained within the framework of the Drude model. According Drude model, the relation between peak position of SPR (λ) and optical dielectric function of the medium ($\varepsilon_m$) can be written as[42]

$$\lambda^2 = \lambda_P^2(\varepsilon^\alpha + 2\varepsilon_m),$$

where $\lambda_P$ and $\varepsilon^\alpha$ are the bulk metal plasmon wavelength and high frequency dielectric constant due to inter-band and core transitions, respectively. Therefore, the graph between $\lambda^2/1000$ versus $2\varepsilon_m$ are plotted (figure 16). Non-linearity is seen in polar solvents but straight line in non-polar solvents, which indicates that µ affects the position of SPR in the frame-work of Drude model only in nonpolar solvents, not in polar solvents[42]. Therefore, it is inferred that SPR is highly sensitive to µ and polarity that shows the extended features.

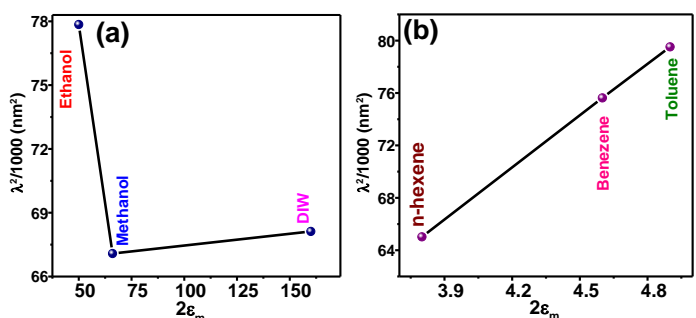

**Figure 16** $\lambda^2/1000$ versus $\varepsilon_m$ curve of (a) polar solvents and (b) non-polar solvents.

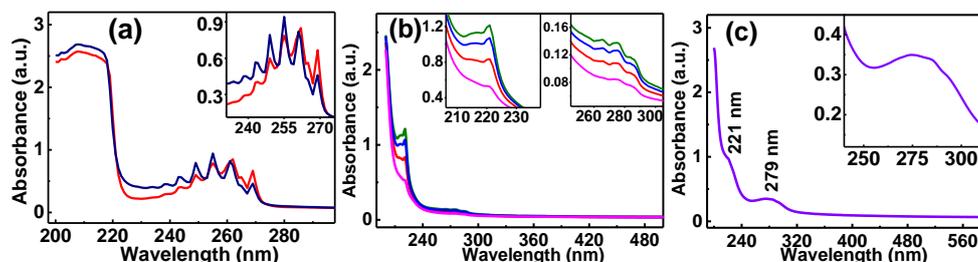

**Figure 17** UV-Visible spectra of Ag4 with varying concentration in (a) n-hexane and (b) ethanol after dispersed using probe-ultrasonication and (c) in ethanol dispersed using magnetic stirrer.

Further, to check our hypothesis of peak splitting, we performed UV-Visible measurements with varying concentration of NPs, ultrasonication time and amplitude in ethanol and n-hexane. UV-Visible spectra of Ag4 were recorded by varying the concentration of NPs using their dilution followed by probe-ultrasonication (of frequency 20 KHz) for 5 min in each time (figure 17). As concentration of NPs increases, SPR gets narrower with enhanced absorbance and gradually blue-shifted. Five (two) peaks can be clearly seen, while two (one) peaks are feeble in ethanol (n-hexane) for 5 min ultrasonication. These feeble intensity peaks increase with increase in concentration and time of ultrasonication (figure 17 (a, b), insets). Now, absorbance spectra of Ag4 in ethanol were taken for 2 min and 80 % less amplitude. A hump like feature (P2) without any splitting is found, which is red-shifted. Lesser amplitude and time of ultrasonication may lead to reduce dispersity of NPs, decrease interparticle distance, weakening the bonding between NPs and solvents molecules, and hence charge transfer effects. Therefore, we observed the tunable features of SPR in Ag NPs in broad range of ultraviolet (UV) regime, which indicates their many potential applications such as photocatalyst, UV-detector and plasmonic sensor[37, 43,44].

**Conclusion**

The SPR of Ag NPs are found to be sensitive to particle size, dielectric environment and solvent refractive index. The main SPR feature shifts towards higher wavelength side (red shift), as particle size decreases or adsorbate coverage increases. The solvent refractive (µ) of methanol, deionized water, ethanol, n-hexane, benzene and toluene has been used to vary µ. The main SPR feature is shifted towards higher wavelength side (red-shift) as µ increases in polar as well as non-polar solvents. We found that SPR near 320 nm in smaller NPs in air shifts to around 259 nm, 261 nm, 277 nm, 255 nm, 275 nm and 282 nm in methanol, deionized water, ethanol, n-hexane, benzene and toluene, respectively. Furthermore, the agglomeration condition studies of the NPs in these solvents using dynamic light scattering through zeta-potential (ζ), hydrodynamic diameter (HD), mobility and conductivity for their correlation with SPR well-support the interactions that are taking place among the NPs and solvents. This has further been substantiated using a model for re-adjustment of the Fermi level of Ag NP and HOMO and LUMO of surface-bound solvents. Finally, the multiple peak features of SPR in ultraviolet region indicate their potential applications such as photocatalyst, UV-sensor and detector.

**Acknowledgements**
Authors would like to acknowledge Dr. D. M. Phase/ Mr. Gyanendra Panchal, Dr. R. J. Choudhary and Dr. U. Deshpande/Mr. Prakash Behera, UGC-DAE Consortium for Scientific Research, Indore, India for providing XRD, XPS and FTIR and UV-Visible reflectance data, respectively. Authors are thankful to USIF, Aligarh Muslim University, Aligarh and CIL, Dr. Harisingh Gour University, Sagar and Dr. Subha Jain, Vikram University, Ujjain, India for providing HRTEM, FESEM and UV-Visible absorbance data, respectively.

# Electronic Supplementary Information

# Influence of surfactant, particle size and dispersion medium on surface plasmon resonance of silver nanoparticles

Vikash Sharma[1,*], Divya Verma[2], Gunadhor Singh Okram[1,#]

[1]UGC-DAE Consortium for Scientific Research, University Campus, Khandwa Road, Indore 452001, Madhya Pradesh, India.

[2] School of Studies in Chemistry & Biochemistry, Vikram University, Ujjain-456010, Madhya Pradesh, India

Email: [#]okram@csr.res.in, [*]vikashphy1@gmail.com

**Table S1** Zeta-potential (ζ), mobility, conductivity of first run along with refractive index obtained from dynamic light scattering measurements of Ag4 in various solvents at pH=7.

| Solvent | Zeta potential (mV) | Mobility (μcm$^2$/Vs) | Conductivity (mS/cm) | Refractive index |
|---|---|---|---|---|
| Methanol | 14.60 | 2.98×10$^{-5}$ | 0.0176 | 1.3270 |
| DIW | -37.23 | -26.25×10$^{-5}$ | 0.0128 | 1.3325 |
| Ethanol | 10.40 | 2.41×10$^{-5}$ | 0.0096 | 1.3617 |
| n-Hexane | 26.94 | 1.34×10$^{-5}$ | 0.0062 | 1.3758 |
| Benzene | 8.70 | 0.38×10$^{-5}$ | 0.0057 | 1.4957 |
| Toluene | 5.10 | 0.37×10$^{-5}$ | 0.0053 | 1.4964 |

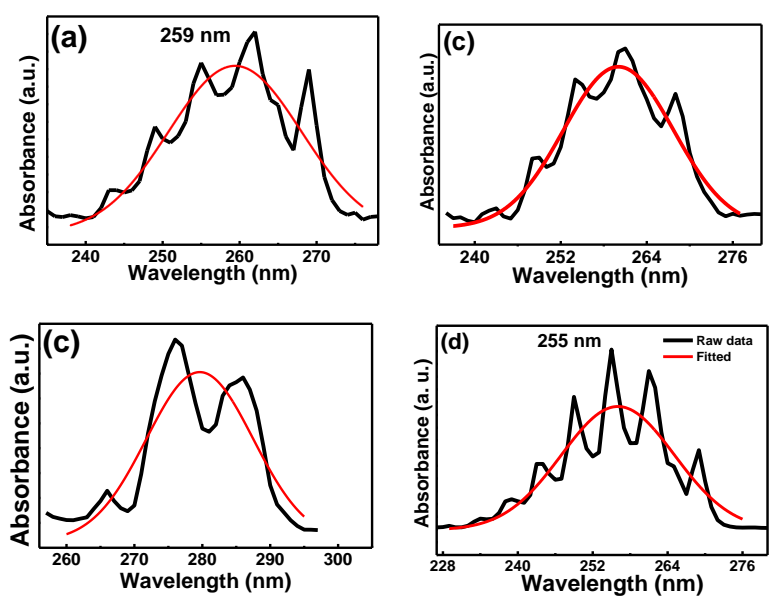

**Figure S1** SPR peak fitting using Gaussian profile after baseline correction of Ag4 in (a) methanol, (b) deionized water, (c) ethanol and (d) n-hexene.